\DocumentMetadata{}
\documentclass[sigconf, screen]{acmart}
\acmConference[FSE 2025]{The ACM International Conference on the Foundations of Software Engineering}{23-27 June, 2025}{Trondheim, Norway}

\AtBeginDocument{%
  \providecommand\BibTeX{{%
    \normalfont B\kern-0.5em{\scshape i\kern-0.25em b}\kern-0.8em\TeX}}}

\setcopyright{acmcopyright}
\copyrightyear{2018}
\acmYear{2018}
\acmDOI{10.1145/1122445.1122456}

\acmConference[FSE '25]{FSE '25: ACM Symposium on the Foundations of Software Engineering}{June 23--27, 2025}{Trondheim, Norway}
\acmPrice{15.00}
\acmISBN{978-1-4503-XXXX-X/18/06}

\acmBooktitle{Companion Proceedings of the 33rd ACM Symposium on the Foundations of Software Engineering (FSE '25), June 23--27, 2025, Trondheim, Norway}
\setcopyright{none}
\renewcommand\footnotetextcopyrightpermission[1]{} % removes footnote with conference information in first column

\usepackage{xcolor}
\usepackage{color}
\usepackage{xspace}
\usepackage{cleveref}
\usepackage[many]{tcolorbox}

\DeclareRobustCommand{\mybox}[2][gray!20]{%
\begin{tcolorbox}[
        breakable,
        left=0pt,
        right=0pt,
        top=0pt,
        bottom=0pt,
        colback=#1,
        colframe=#1,
        width=\linewidth, 
        enlarge left by=0mm,
        boxsep=5pt,
        arc=0pt,outer arc=0pt,
        ]
        #2
\end{tcolorbox}
}

\newcommand{\tool}{LiveSCA\xspace}

\author{
Lyuye Zhang$^{1}$,
Chengwei Liu$^{1}$*,
Jiahui Wu$^{1}$,
Shiyang Zhang$^{2}$,
Chengyue Liu$^{1}$,
Zhengzi Xu$^{3}$,\\
Sen Chen$^{4}$,
Yang Liu$^{1}$\\
\small{$^{1}$College of Computing and Data Science, Nanyang Technological University, Singapore \\
$^{2}$College of Intelligence and Computing, Tianjin University, China \\
$^{3}$Imperial Global College, Singapore \\
$^{4}$College of Cryptology and Cyber Science, Nankai University, China \\
}
}
\thanks{*Chengwei Liu is the corresponding author.}
\begin{document}

\title{Drop the Golden Apples: Identifying Third-Party Reuse by DB-Less Software Composition Analysis}

\renewcommand{\shortauthors}{Zhang, et al.}

\begin{abstract}
The prevalent use of third-party libraries (TPLs) in modern software development introduces significant security and compliance risks, necessitating the implementation of Software Composition Analysis (SCA) to manage these threats.
However, the accuracy of SCA tools heavily relies on the quality of the integrated feature database to cross-reference with user projects. 
While under the circumstance of the exponentially growing of open-source ecosystems and the integration of large models into software development, it becomes even more challenging to maintain a comprehensive feature database for potential TPLs.
To this end, after referring to the evolution of LLM applications in terms of external data interactions, we propose the first framework of DB-Less SCA, to get rid of the traditional heavy database and embrace the flexibility of LLMs to mimic the manual analysis of security analysts to retrieve identical evidence and confirm the identity of TPLs by supportive information from the open Internet.
Our experiments on two typical scenarios, native library identification for Android and copy-based TPL reuse for C/C++, especially on artifacts that are not that underappreciated, have demonstrated the favorable future for implementing database-less strategies in SCA.

\end{abstract}


% \ccsdesc[500]{Computer systems organization~Embedded systems}
% \ccsdesc[300]{Computer systems organization~Redundancy}
% \ccsdesc{Computer systems organization~Robotics}
% \ccsdesc[100]{Networks~Network reliability}

% \keywords{datasets, neural networks, gaze detection, text tagging}

\maketitle

\section{Introduction}
In modern software development, the prevalence of third-party libraries (TPLs) and extensive code reuse have introduced significant risks in terms of security vulnerabilities and compliance issues. 
To address these challenges, Software Composition Analysis (SCA) has emerged as a vital practice. 
SCA tools systematically identify TPLs within user projects and cross-reference these elements with databases of known security threats and licensing conflicts.
Popular SCA tools, such as Snyk~\cite{snyk}, Synopsys Black Duck~\cite{BlackDuck}, and WhiteSource~\cite{Mendio} have gained wide adoption due to their effectiveness in securing software supply chains. 

Many researchers~\cite{zhan2021atvhunter, zhang2018detecting, backes2016reliable, li2017libd, woo2021centris, wu2023ossfp} from both academia and industry have investigated a lot on the improvement and adoption of SCA tools by different means for various scenarios. 
However, most existing works only validate their SCA tools by prototype databases, for instance, being constructed based on only tens of thousands of repositories with over 100 stars~\cite{woo2021centris,wu2023ossfp,woo2023v1scan}. They reported that the completeness of their component database has a significant negative impact on the accuracy of their tools.
While in real-world scenarios, the candidates of third-party reuse are way more than those included in experimental databases, which makes it difficult to fully validate the effectiveness of proposed SCA tools. 
% Moreover, the effectiveness of SCA tools hinges significantly on the comprehensiveness and accuracy of their underlying databases. 
This issue is particularly critical for C/C++ libraries due to the absence of a centralized package repository and enormous scattered libraries over the Internet reported by Tang et al.~\cite{tang2022towards} unlike the centralized repository PyPI~\cite{pypi} for Python packages. Consequently, accurately mapping detected third-party code to previously collected libraries becomes a significant challenge, complicating the management and maintenance of open-source software.

However, it is also non-trivial to construct a comprehensive database for SCA. Specifically, under the circumstance of the exponentially growth of open source ecosystems, and especially as large models and frameworks gradually permeate the software development, maintaining a comprehensive and reliable feature database is becoming increasingly burdensome for SCA tools. 
Existing commercial SCA tools~\cite{snyk,BlackDuck,Sonatype,OWASP,dependabot} rely heavily on continuously updated databases to maintain seemingly up-to-date collections of libraries. However, this approach is largely confined to well-maintained and centralized repositories such as PyPI, and often overlooks the vast number of self-hosted or less-visible libraries distributed across the Internet. As a result, these tools may operate on incomplete library datasets, limiting their effectiveness in comprehensive dependency detection and vulnerability analysis.
% To this end, it is urgently required to revisit the current SCA framework and follow-up tasks to figure out new solutions for third-party auditing.

Meanwhile, the wide adoption of LLMs~\cite{chang2024survey} inspired us to think from a different angle. Reviewing the evolving interaction ways of LLMs to external data, it turns from heavy solutions, like fine-tunes based on a well-constructed and specialized datasets, to relatively light-weighted solutions, such as retrieval-augmented generation (RAG), which integrates small but more focused datasets by relying more on the generalizability of LLMs, and to more flexible solutions, such as LLM agents that can directly integrate searching engines based on the reasoning capability of LLMs into their task orchestration. 

To this end, we first propose the concept of DB-Less SCA. Unlike traditional SCA tools that require a comprehensive database for cross-referencing, instead, we switch the identification of TPLs to the process of how security analysts manually figure out the correct identity of TPLs, by incorporating the manual process substituted by agents, with a well-orchestrated chain of orders based on LLMs. 

However, to get rid of the heavy database and propose an actionable solution for DB-Less SCA, we still face the following challenges:
\textbf{C1) Inadequate Identity Evidence.} Unlike traditional SCA tools that match features that are well prepared in databases as evidence of identity, manual analysis follows completely different strategies and clues, based on experiences, to identify TPLs, it is non-trivial to systematically collect these potential evidence of identities.
\textbf{C2) Comprehensive Information Searching.} After collecting identity evidence, analysts usually search and find supportive documentation or claims on information sources that are relevant to this evidence and can infer the original sources of suspected TPLs. To this end, it is also important to learn insights from analysts and orchestrate the pipelines to precisely pinpoint necessary supportive materials.
\textbf{C3) Insufficient Validation.} Another major concern lies in what extent to which the supportive documentation and claims can confirm the originality of TPLs. Therefore, it also matters to properly introduce the knowledge of analysts to the automated judging and reasoning of the rationale of the originality of TPLs.

Therefore, we propose the first \underline{live} Database-less \underline{S}oftware \\
\underline{C}omposition \underline{A}nalysis (\tool) with live and real-time data across the Internet, which incorporates LLMs and human expertise together to re-orchestrate the standardized process of software composition analysis for TPLs. 
To address C1, we have developed a general evidence-collection step that focuses on capturing the textual semantics of the target library. This process goes beyond the low-level semantics, such as code structure, typically extracted by traditional SCA tools. Instead, we target textual semantics that describe the library's functionality, enabling the subsequent LLM agents to grasp the essence of the library.
For C2, we utilize a multi-agent framework that iteratively scrapes, comprehends, and compares Internet search results. This synergistic approach aims to identify the most promising target webpage based on previously collected evidence. The process is designed to mimic human manual checking, performed efficiently by multiple LLM agents.
Regarding C3, we employ another agent specifically for independent validation. This agent not only validates the findings but also feeds back its determinations to the evidence summarization agent. This feedback mechanism dynamically adjusts the search objectives, enhancing the accuracy and relevance of the search process.

Based on the evaluation, \tool achieved success rates of 59.20\% and 57.50\% for two typical SCA applications.
Though not comparable with existing traditional SCA tools with high-quality databases, these results still demonstrate the promising potential of our solution in the context of DB-less SCA, especially for TPLs that are not underappreciated by downstream users. The outcomes suggest a favorable future for implementing database-less strategies in SCA, indicating that even without traditional databases, effective library detection and verification are achievable.

% \section{Background}

\section{Related Work of Traditional SCA Tools}
Both industry and academic tools considered state-of-the-art perform Software Composition Analysis (SCA) through either Software Bill of Materials (BOM)~\cite{SBOM2022} detection or code clone mappings.
% When formatted and official file listed as the TPL dependency list on the target project, BOM detecting tools can thoroughly list all TPLs used in the project with all the detailed vendor, name, and version.
When BOMs, formatted as official files, list the TPLs dependencies of a project, BOM detection tools can comprehensively enumerate all TPLs used, including detailed information about the vendor, name, and version.
% Tools such as OWASP~\cite{OWASP} and Sonatype~\cite{Sonatype} rely on the BOM files to match the listed names on the SBOM files with the names of pre-collected TPLs in the database.
Tools like OWASP~\cite{OWASP} and Sonatype~\cite{Sonatype} utilize BOM files to align the names listed in Software Bills of Materials (SBOM) with those of TPLs previously collected in a database.
% However, this approach is effective mainly for those programming languages that has official package manage which maintain a complete list of all the TPLs, since it rely on the the BOM file formatted by the package manages.
However, this method is mainly effective for programming languages that have an official package manager, which maintains a comprehensive list of TPLs, as it depends on the BOM files formatted by these package managers.
% However, for C/C++, there is no official or widely adopted package manager that can maintain a complete list of TPLs.
For C/C++, no official or universally adopted package manager exists to maintain a complete list of TPLs.
% The state-of-the-art C/C++ TPL detection tools are thus adopting the code clone mapping between the target project and pre-constructed database.
Consequently, leading C/C++ TPL detection tools adopt code clone mapping strategies between the target project and a pre-established database.
% FOR EXAMPLE, CENTRIS, OSSFP, LIBDB....
% BlackDuck~\cite{BlackDuck} and Snyk CLI~\cite{snky}, ATVHunter~\cite{zhan2021atvhunter}, LibPecker~\cite{zhang2018detecting}, LibScout~\cite{backes2016reliable}, and LibD~\cite{li2017libd} use code features such as control flow graphs to conduct code clone mappings between the feature database and the target projects.
For instance, tools like BlackDuck~\cite{BlackDuck}, Snyk CLI~\cite{snyk}, ATVHunter~\cite{zhan2021atvhunter}, LibPecker~\cite{zhang2018detecting}, LibScout~\cite{backes2016reliable}, and LibD~\cite{li2017libd} utilize code features like control flow graphs to perform code clone mappings between the feature database and target projects.
% Tools like CENTRIS~\cite{woo2021centris}, OSSFP~\cite{wu2023ossfp} improve the accuracy by selecting more representative feature extracted from the GitHub repositories over 100 star gazes.
Tools such as CENTRIS~\cite{woo2021centris} and OSSFP~\cite{wu2023ossfp} enhance accuracy by selecting more representative features extracted from GitHub repositories with over 100 stars.
% None of these state-of-the-art tools use a complete list of the TPLs to construct their feature database, which leads to high potential false negative cases.
None of these advanced tools utilize a complete list of TPLs to construct their feature databases, potentially leading to a high incidence of false negatives.

% However, collecting the complete list of the C/C++ TPLs is an extremely difficult task for several reason. 
% First, thes is no existing official pacakge manager for all the TPL vendor or author to formally register their libraries.
% Second, the C/C++ TPLs are not all maintained in any single common platform, some of which are even hosted on their personal website or specific homepage.
% These situations make  collecting a complete list of TPLs  an  unrealistic path for address the limitation of the state-of-the-art C/C++ TPL detection tools.
% However, all the SCA tools cannot performed without the feature database which the completeness would directly affect their accuracy.
% Thus, a alternative way to achieve better performance is using the DB-Less method.
% -------------------------------------- GPT rewrite --------------------------------------
% However, collecting the complete list of the C/C++ TPLs is an extremely difficult task for several reason.
However, compiling a complete list of C/C++ TPLs is exceptionally challenging for several reasons.
% First, thes is no existing official pacakge manager for all the TPL vendor or author to formally register their libraries.
First, there is no existing official package manager where all TPL vendors or authors can formally register their libraries.
% Second, the C/C++ TPLs are not all maintained in any single common platform, some of which are even hosted on their personal website or specific homepage.
Second, C/C++ TPLs are not maintained on a unified platform; instead, some are hosted on personal websites or specific homepages.
% These situations make  collecting a complete list of TPLs  an  unrealistic path for address the limitation of the state-of-the-art C/C++ TPL detection tools.
These circumstances render the task of compiling a comprehensive list of TPLs an impractical approach to addressing the limitations of current C/C++ TPL detection tools.
% However, all the SCA tools cannot performed without the feature database which the completeness would directly affect their accuracy.
Moreover, all Software Composition Analysis (SCA) tools rely on a feature database, the completeness of which directly impacts their accuracy.
% Thus, a alternative way to achieve better performance is using the DB-Less method.
% Therefore, an alternative approach to enhancing performance is to employ the database-less (DB-Less) method.
Our database-less approach represents a significant departure from traditional SCA methodologies, which have historically depended on extensive and often cumbersome databases. 
Surprisingly, despite the obvious advantages of reducing reliance on heavy databases, this innovative strategy has not yet been considered or explored within both industry and academia.

\begin{figure*}[t!]
	\centering
	\includegraphics[width=0.9\linewidth]{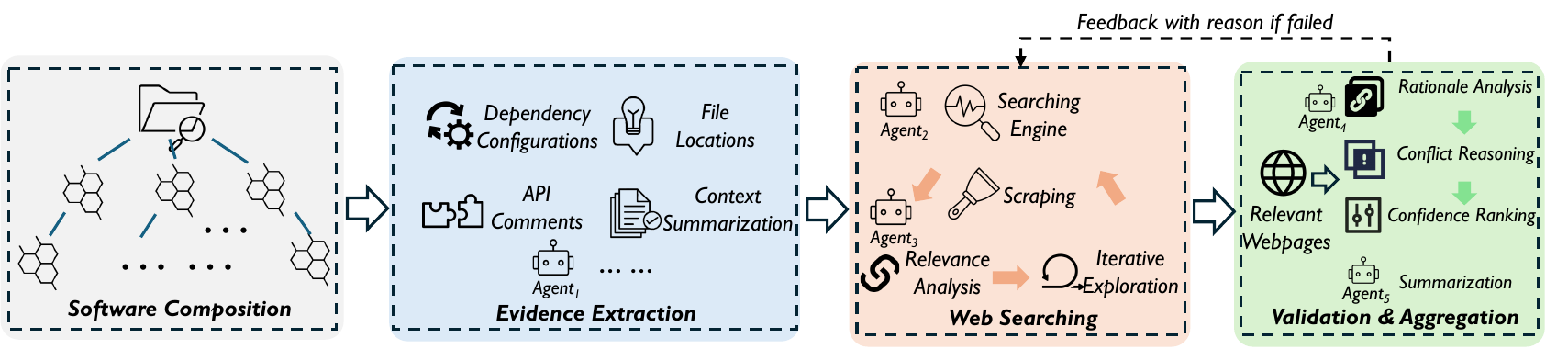}
	\caption{Overview of \tool}
	\label{fig:overview}
\end{figure*}

\section{The Live DB-Less SCA Framework}
\tool aims to facilitate a database-less SCA, eliminating the dependence on any pre-established databases. As indicated in \Cref{fig:overview}, the methodology employed by \tool capitalizes on a multi-agent framework to autonomously analyze potential open-source libraries and localize their origins. For clarity of notation, each agent is denoted as $A_{n}$, where $n$ represents the agent’s unique identifier. Unless otherwise specified, all agents operate independently, each implemented with a separate LLM model session to perform tasks autonomously. It leverages Internet-based knowledge retrieval to ascertain the identities and origins of these libraries. Specifically, \tool initially extracts and analyzes evidence from target projects. Subsequently, it heuristically summarizes the core descriptions of potential libraries based on the evidence. Thereafter, \tool engages in an iterative process to crawl and analyze webpages to trace their origins. Ultimately, \tool ranks the filtered webpages and validates the derived origins.

Given the diverse scenarios involved in SCA for TPL identification, we selected two typical scenarios to demonstrate \tool: identifying native libraries in Android application packages (APKs)~\cite{almanee2021too} and detecting cloned TPLs in C/C++ projects~\cite{woo2021centris}. The identification of native libraries in Android apps is particularly crucial because these libraries, often written in C/C++, are commonly utilized as third-party components without uniform package management. It often results in insufficient evidence regarding how these libraries are compiled and incorporated, leading to challenges, such as vulnerabilities inherent in these libraries, posing security risks to the Android application. 
% Consequently, the first scenario selected for in-depth discussion is the identification of native libraries in Android apps.
The second scenario is the identification of cloned source code in C/C++ libraries. Due to the absence of a uniform package manager for C/C++, libraries are often directly cloned into projects unlike other programming languages like Java. Therefore, it is crucial to identify the cloned code in C/C++ projects and their sources for subsequent analysis concerning security and licensing. Contemporary tools such as Centris~\cite{woo2021centris} and OSSFP~\cite{wu2023ossfp} are designed to extract precise features for library identification. However, these tools depend on an existing database that incorporates features from previously identified libraries, which can be overcome by \tool.
% Lacking a complete database, these tools struggle to accurately identify libraries. 
% Consequently, we propose using LLMs and the Internet to overcome these limitations in library identification.

\subsection{Evidence Extraction}
The initial step in the SCA process involves extracting features from the target projects. Notably, SCA tools may accept various formats, such as source code, bytecode, or binary files. Thus, this extraction step is tailored to specific software formats. Regardless of the formats, the fundamental objective remains the same: to gather evidence that aids in the identification of libraries and the retrieval of their origins. This phase also involves filtering out extraneous noise, such as excessive strings extracted from ELF files. Given the dependency on format-specific procedures, detailed implementations are discussed in the Implementation~\Cref{sec:impl}. The general evidence includes dependency configurations, file locations, API comments or documentation, and the relevant code context, but the availability depends on specific tasks.
% The noise reduction process is also dependent on the software format.

% Feature extraction in our DB-less SCA deviates from traditional SCA methods, which typically rely on a feature database to compare library signatures for identification. Traditional SCA targets efficient comparisons through abstract signature calculations, facilitating effective library matching. In contrast, our approach leverages the reasoning and analytical capabilities of LLMs, primarily focusing on the text-based semantics of target libraries, rather than utilizing hash calculations from the source code, which are common in traditional SCA. \tool primarily extracts readable features, enhancing human understanding and supporting an explainable reasoning process. .

\subsection{Web Searching}
The subsequent step involves searching for the origins of the target libraries, such as their homepages and repositories. This process is initiated with inputs derived from the previously extracted evidence. \tool employs a multi-agent strategy to heuristically analyze evidence and dynamically adjust the search process. Specifically, three agents are deployed for optimized searching. Agent $A_1$ processes the evidence to identify relevant keywords that encapsulate the core functionalities of the libraries. Based on these keywords, $A_1$ retrieves and saves the first two pages of results from Google, including titles, snippets, and URLs. For example, the keywords for Glad were \textit{glad, OpenGL, loader, library, and CMake}.

Agent $A_2$ then collects the search results and iteratively scrapes the web pages by Scrapy~\cite{scrapy} into texts only. Considering that webpage contents can be extensive, $A_2$ is designed to synthesize and summarize each page given the page content from the scraping, producing a concise summary that highlights the primary uses of the webpage. Consequently, each webpage is tagged with a brief summary and, as a fallback, a snippet from Google if the webpage crawling does not succeed.
An example of Glad\footnote[1]{https://github.com/Dav1dde/glad} is \textit{"GLAD is a multi-language Vulkan/GL/GLES/EGL/GLX/WGL loader-generator based on official OpenGL specifications. It is commonly used to load OpenGL functions and extensions in graphics programming."}

Lastly, Agent $A_3$ evaluates the correlation between the evidence of the target library and the information on the webpages, focusing particularly on determining the originality of the sources. This is critical because popular libraries often have numerous discussions across forums and multiple forked repositories that are not the original source. $A_3$ ranks the webpages based on this correlation according to its own reasoning process given the webpage summaries, aiming to identify and list the most authentic and relevant sources. For the Glad case, the homepage\footnotemark[1] is found and ranked at the first place.

% As \tool independently searches for the source code webpage and the homepage, it employs concurrent pipelines to manage these tasks separately. This parallel processing allows \tool to search and compile ranked lists for both the source code and the homepage independently. Upon completing these searches, \tool returns these lists for aggregation in the subsequent step.

\subsection{Validation and Aggregation}
The final major step in \tool involves aggregation and dynamic validation, orchestrated by two specialized agents. The validation process, managed by $A_4$, incorporates a feedback mechanism that may refine the extracted evidence. To ensure integrity, $A_4$ operates independently from the earlier agents, avoiding shared knowledge that could bias the validation outcomes. $A_4$ assesses the alignment between the top-ranked webpages against the initially extracted features. If misalignments are determined by the LLM output, the reasons generated by $A_4$ are then fed back to the evidence summarization module to refine the associated keywords. This validation feedback loop is executed three times to optimize the search results.
Upon validation, if $A_4$ confirms the top results as the target homepage or the loop limit is reached, the information is relayed to $A_5$ for aggregation. $A_5$ then compiles the data, generates a brief description of the target library, and returns this packaged metadata for the libraries, which could be listed in the SBOM.
% To prevent duplicate entries, \tool checks for existing libraries in the database with the same name. $A_5$ compares the new descriptions against existing entries to ensure no duplicates are saved in the database.

\subsection{Implementation}
\label{sec:impl}
\tool is implemented using the widely adopted multi-agent framework, Dify~\cite{dify}, which orchestrates workflows based on LLMs with flexible, customized tools. The entire reproducible package of Dify DSL is packed and stored online \footnote{https://drive.google.com/file/d/1BsbohzZjfCkw2VQIhWCsuFmh6EAie4eM/} with detailed prompts.
% This enables the execution of complex tasks through multi-agent collaboration. 
Within this framework, agents are implemented using GPT-4o~\cite{gpt4o}, the state-of-the-art (SOTA) model renowned for its efficiency in handling a variety of text-based tasks.

For specific scenarios, the evidence extraction of \tool varies significantly, necessitating customized tools tailored to software formats. In the context of native library identification in Android applications, \tool initiates the process by extracting the APK~\cite{Android2020} files and examining the ELF~\cite{ELFSpec1995} files in SO~\cite{Rosen2007} format contained within. Recognizing that the file name of an SO file often hints at the native library name, \tool initially saves this file name.
Following this, \tool employs the \texttt{strings} command to extract readable strings from the SO file. Given the variability in how vendor or version information is represented across different SO files, \tool selectively filters these strings based on their length (shorter than 10 characters are discarded), as they typically lack meaningful information. 
% The remaining strings are then compiled for further analysis.

For TPL recognition in C/C++ source projects, \tool leverages on the file structure to pinpoint potential TPL locations. Utilizing the \texttt{tree} command, \tool extracts the file structure of the target projects and inputs this data into an LLM agent. This agent then identifies potential TPLs and provides a list of the top five files deemed worth inspecting.
Subsequently, \tool reads the content of these specified files for each TPL. The contents of these files are extracted and used as preliminary evidence for further processing. 

% This method ensures a targeted and efficient approach to identifying and analyzing TPLs within complex project structures.

% The remaining steps in the SCA process are uniformly applied across various tasks and scenarios. For these steps, the LLM agents employed utilize GPT-4o models to ensure optimized performance. Additionally, the web scraping tasks are facilitated by the built-in scraper of Dify, which is integrated to enhance efficiency and streamline the data acquisition process.

\section{Experiments}
In our experimental setup, we employ two typical SCA scenarios to demonstrate the efficacy of our framework, \tool. This experiment is designed to address two pivotal research questions:
% \begin{itemize}[leftmargin=9pt]

    \noindent $\bullet$ RQ1: How effectively does \tool identify native libraries within Android applications?
    
    \noindent $\bullet$ RQ2: How effectively does \tool recognize open-source libraries in C/C++ projects?
% \end{itemize}

\subsection{Dataset Collection}
For RQ1, our dataset collection involved Android applications from Google Play. Initially, we gathered the top 100 applications and extracted their corresponding SO files, resulting in a total of 11,205 files with duplicates. These files were associated with 108 different libraries, where various versions corresponded to distinct SO files. To validate the effectiveness of \tool, we randomly selected one SO file from each library, yielding a sample of 108 SO files. 
% Each file was then manually verified to confirm its source. 
% With this dataset, \tool will now commence tracing the origins of these SO files, leveraging only the SO file data.

For RQ2,
% to establish a reliable ground truth for evaluating the accuracy of the SCA tools in our research, 
we began by gathering a dataset of C/C++ projects from GitHub
% Utilizing searching interface provided by GitHub , we 
by filtering for projects that not only had over 100 stars but were primarily developed in C/C++, yielding 23,568 projects.
For manual confirmation, we randomly inspected projects and manually confirmed the usage of TPLs of them until the projects reached 100. Then, we randomly selected 200 confirmed TPL as the dataset.
% To further refine our dataset to suit the specific needs of our study, we randomly selected 100 projects from this pool, with an additional requirement: each project must contain a dedicated folder labeled "3rdParty" or similar, indicating the use of third-party libraries. 
% This selection criterion was critical as it ensured that the chosen projects were appropriate for evaluating the effectiveness of SCA tools in identifying TPLs.

\subsection{RQ1: Native library Identification}

% Please add the following required packages to your document preamble:
% \usepackage{booktabs}
\begin{table}[]
\label{tab:rq1}
\caption{Effectiveness of \tool for Typical SCA Scenarios}
\begin{tabular}{@{}lrrrr@{}}
\toprule
Scenario    & \#Total & \#Collected URL & \#Correct & \#Hints Found \\ \midrule
SO in APK   & 108     & 64              & 64        & 102           \\
Clone C/C++ & 200     & 119             & 105       & 121           \\ \bottomrule
\end{tabular}

\end{table}

In the RQ1, we aimed to identify the sources, including homepages and repositories, of 108 native libraries based solely on given SO files. The results show that out of the 108 libraries, 64 (59.25\%) were correctly located. A manual analysis of the URLs of the collected web pages validated the accuracy by the first three authors. 
% validating the effectiveness of the separate agent tasked with ensuring that the output links meet the specified requirements.

Further analysis was conducted on the unsuccessful cases. \tool encountered network failures in 6 instances, leading to aborted processes. Excluding these, \tool returned results for 102 cases, with 64 successful identifications included. For the 38 cases where sources could not be located, \tool had correctly collected vendor information, but no accessible URLs were available. Specifically, 26 of these cases involved source URLs that led to \texttt{404} errors, indicating discontinued maintenance. Additionally, manual searches for the remaining 12 cases did not yield any results, underscoring the limitations of relying solely on Google search capabilities.
Even in the 38 cases where sources could not be located, \tool successfully returned vendor information deduced from the extracted strings. This information can still be leveraged to manually locate additional related webpages or undertake further searches.
% The availability of vendor details, despite the absence of direct URLs, provides valuable leads that could aid in deeper investigations or future enhancements of our framework's searching algorithms. 
Thus, even the unsuccessful cases contribute valuable data for refining \tool's capabilities and for potentially guiding manual follow-up searches or other fine-grained tools to uncover sources that are not readily accessible via \tool.

Overall, this task took an average time of 27.99s with 10,196.85 tokens with an estimated cost of less than 3 USD.

\mybox{
\textbf{Answering RQ1}: \tool successfully located the sources for 64 (59.20\%) out of 108 native libraries. Additionally, it was able to provide hints about vendors and version information in 102 cases. These hints could potentially be leveraged for further manual searches or to enhance automated querying techniques. 
% This demonstrates \tool's effectiveness in accurately identifying sources for native C/C++ libraries.
}

\subsection{RQ2: Cloned TPL in C/C++ Projects}

In the RQ2 subsection, we randomly sampled 200 distinct cases to verify \tool's performance. As indicated in~\Cref{tab:rq1}, out of these cases, \tool successfully identified the sources for 115 cases (57.50\%). However, it initially found sources for 119 TPLs, but 4 of these were inaccurately identified upon manual verification. These inaccuracies primarily stemmed from the broad or ambiguous meanings associated with certain library names, which could refer to multiple different entities. Common issues included overly short names or names that overlapped with common terms. For example, the incorrect TPL homepage for \textit{semver.com} was identified because the library name was \texttt{semver}, which, in the specific context, referred to a C/C++ library implementing Semantic Versioning. In contrast, \textit{semver.com} hosts general documentation about Semantic Versioning rules, not a specific library.

This error highlights a challenge in using LLM agents: distinguishing between closely related concepts that have nuanced differences can be difficult. \textit{semver.com} was flagged by the LLM due to its strong relevance to Semantic Versioning, albeit not as a direct source for the C/C++ library. Upon further manual investigation, the correct URL for the C/C++ library was identified as \textit{https://github.com/h2non/semver.c}. This case exemplifies the need for enhanced understanding in LLMs to differentiate between contextually similar but distinct entities.

% % Please add the following required packages to your document preamble:
% % \usepackage{booktabs}
% \begin{table}[]
% \caption{Effectiveness of \tool for C/C++ TPL recognition}
% \begin{tabular}{@{}lrrrr@{}}
% \toprule
% Name                 & \#Total & \#Collected URL & \#Correct URL & \#Hints Found \\ \midrule
% \tool & 200     & 119             & 105       & 121           \\ \bottomrule
% \end{tabular}
% \label{tab:rq2}
% \end{table}

We conducted a detailed analysis of the LLM agent's reasoning process for the cases where no URL was returned, identifying three primary reasons for these failures: (1)The source code files' extracted contents lacked distinctive features necessary for identifying the origin of the libraries. This issue was primarily due to incorrect enumerations in the target files during the feature extraction step, resulting in unclear keywords that led to irrelevant search results. (2) The websites identified through the search were no longer accessible, rendering the URLs unusable.
(3)Ambiguous keywords and generic library names often retrieved a plethora of unrelated web pages, obscuring the target sources and leading to unsuccessful searches.

The average time \tool took to complete the dataset is 29.36s with 15,166.62 tokens with an estimated cost of less than 8 USD.

\mybox{
\textbf{Answering RQ2}: \tool correctly found sources for 115 out of 200 C/C++ TPLs from the C/C++ projects. This demonstrates a substantial success rate of 57.5\%, highlighting the tool's effectiveness in accurately locating library sources.
% The results reflect the capability of \tool to leverage sophisticated LLM techniques for distinguishing and verifying the origins of TPLs. 
% Despite these successes, the experiments also revealed areas for improvement. We identified common issues that led to unsuccessful cases, including insufficiently distinctive features extracted from the source code, inaccessible URLs, and the retrieval of numerous unrelated web pages due to ambiguous keywords. Each of these challenges presents an opportunity for further refinement of \tool's algorithms to enhance its precision and reliability in future deployments.
}

\section{Discussion}

\subsection{Limitations}
As a preliminary prototype, our framework, \tool, exhibits several limitations that impact its effectiveness. The primary limitation lies in the LLM's capacity to handle large text corpora. This constraint may lead to the omission of critical details, adversely affecting overall performance. To mitigate this, we have designed a standalone agent specifically to comprehend and summarize the essence or semantics of the corpus for further processing. 
% Although this approach may still experience performance degradation, it provides a feasible solution for managing excessively long texts.

The second limitation concerns \tool's reliance solely on Internet searches, restricting the scope to publicly available information. Private homepages, repositories, or data not currently available on the Internet remain inaccessible. Extending the search scope to include non-public data sources and historical webpage snapshots could enhance the tool's performance.

A third limitation is that our DB-less SCA framework primarily focuses on coarse-grained TPL source pinpointing. Fine-grained identification, which necessitates detailed detection of specific versions, is not currently supported due to the reliance on coarse evidence. For more precise identification, a more detailed extraction of library evidence is required to support subtle version differences.

Lastly, the framework lacks a comprehensive end-to-end validation process. Ideally, the DB-less SCA framework should validate collected sources at the semantic level, allowing the returned sources to be fed back into the evidence extraction step. This would enable the extraction of adjusted evidence from the returned sources for cross-verification with initial evidence.
% Currently, as a prototype, \tool has not implemented this feature.

\subsection{Threats to validity}
One of the primary threats to the validity of our work concerns the ground truth labeling process. Due to the necessity of manually reviewing collected webpages to understand the potential usage relationships between the user's project and the identified libraries, the risk of errors in labeling is significant. 
% Mistakes can occur, particularly in cases where the homepage or repository identified may not correspond to the target library actually utilized in the project.
To mitigate this risk, we have implemented a majority voting scheme conducted by the first three authors. This approach aims to minimize subjective biases and errors by ensuring that at least two out of the three authors agree on the relevance and accuracy of each labeled instance. 

Another potential threat lies in the dataset construction, as the current dataset includes only a limited number of samples, which may affect its representativeness. To mitigate this issue, we selected widely used Android applications along with distinct native libraries, as well as popular C/C++ repositories, to ensure diversity and relevance. In future work, we plan to incorporate a more extensive dataset to support a more comprehensive evaluation.

The settings and configurations of the employed techniques may pose a threat to the validity. For instance, the output of Scrapy used for web scraping depends on its specific configuration, which influences the extracted textual content. Similarly, parameters such as the depth level of the \texttt{tree} command and file extension filters can affect the scope and quality of the extracted evidence. Additionally, the temperature setting of GPT-4o may impact generation outcomes. To mitigate these concerns, we adopted widely accepted or recommended configurations. In particular, we set the temperature to zero to ensure consistency and reproducibility of results.

\section{Conclusion}
In conclusion, the development and evaluation of \tool have demonstrated its capability to effectively identify native libraries in Android apps and recognize cloned C/C++ TPLs, achieving success rates of 59.20\% and 57.50\% respectively. These results underscore the viability of a multi-agent, DB-less SCA framework, showcasing its potential to operate efficiently without reliance on traditional databases and overcome the current bottleneck of the incomplete pre-built database.
% The promising outcomes from this preliminary prototype highlight the framework's ability to adapt to complex SCA scenarios through advanced LLM techniques and a dynamic, synergistic approach among multiple agents. Moving forward, further refinement and expansion of the tool’s capabilities could lead to more robust, accurate, and scalable solutions for software composition analysis. The future of DB-less SCA appears promising, with potential applications extending beyond the current scope to broader contexts within software engineering and security.

% \clearpage

\section*{Acknowledgment}
This research is supported by the Ministry of Education, Singapore, under its Academic Research Fund Tier 1 (RG96/23). It is also supported by the National Research Foundation, Singapore, and DSO National Laboratories under the AI Singapore Programme (AISG Award No: AISG2-GC-2023-008); by the National Research Foundation Singapore and the Cyber Security Agency under the National Cybersecurity R\&D Programme (NCRP25-P04-TAICeN) and CyberSG R\&D Cyber Research Programme Office; and by the National Research Foundation, Prime Minister’s Office, Singapore under the Campus for Research Excellence and Technological Enterprise (CREATE) programme.
% This research is supported by the National Research Foundation, Singapore, and Cyber Security Agency of Singapore under its National Cybersecurity R\&D Programme.
Any opinions, findings and conclusions, or recommendations expressed in these materials are those of the author(s) and do not reflect the views of National Research Foundation, Singapore, Cyber Security Agency of Singapore as well as CyberSG R\&D Programme Office, Singapore.

\bibliographystyle{ACM-Reference-Format}
\bibliography{acmart}
\end{document}